RESEARCH ARTICLE       OPEN ∂ ACCESS

# Composite analysis with Monte Carlo methods: an example with cosmic rays and clouds

B.A. Laken[1,2,*] and J. Čalogović[3]

[1] Instituto de Astrofísica de Canarias, Via Lactea s/n, 38205 La Laguna, Tenerife, Spain
   *Corresponding author: blaken@iac.es
[2] Department of Astrophysics, Faculty of Physics, Universidad de La Laguna, 38205 La Laguna, Tenerife, Spain
[3] Hvar Observatory, Faculty of Geodesy, University of Zagreb, Kaciceva 26, HR-10000, Zagreb, Croatia

### ABSTRACT

The composite (superposed epoch) analysis technique has been frequently employed to examine a hypothesized link between solar activity and the Earth's atmosphere, often through an investigation of Forbush decrease (Fd) events (sudden high-magnitude decreases in the flux cosmic rays impinging on the upper-atmosphere lasting up to several days). This technique is useful for isolating low-amplitude signals within data where background variability would otherwise obscure detection. The application of composite analyses to investigate the possible impacts of Fd events involves a statistical examination of time-dependent atmospheric responses to Fds often from aerosol and/or cloud datasets. Despite the publication of numerous results within this field, clear conclusions have yet to be drawn and much ambiguity and disagreement still remain. In this paper, we argue that the conflicting findings of composite studies within this field relate to methodological differences in the manner in which the composites have been constructed and analyzed. Working from an example, we show how a composite may be objectively constructed to maximize signal detection, robustly identify statistical significance, and quantify the lower-limit uncertainty related to hypothesis testing. Additionally, we also demonstrate how a seemingly significant false positive may be obtained from non-significant data by minor alterations to methodological approaches.

## 1. Introduction

The composite analysis technique, also referred to as a superposed epoch analysis, and conditional sampling, is used in numerous fields including geostatistics, fluid dynamics, and plasma physics. It has been frequently employed to examine a hypothesized link between atmospheric properties and sudden decreases in cosmic ray intensity over daily timescales (Forbush decreases). The composite technique is useful for isolating low-amplitude signals within data where background variability would otherwise obscure detection (Chree 1913, 1914). Composite analyses rely on selecting subsets of data, wherein key points in time can be identified based on some criteria, e.g., the occurrence of unusual or extreme events in one or more datasets (Chree 1913, 1914; Forbush et al. 1983). Through the accumulation and averaging of successive key events, the stochastic background variability may be reduced to a point where low-amplitude signals become identifiable.

To exemplify the composite methodology and how it may be applied to isolate low-amplitude signals, we present the following example. Imagine a regularly sampled time series, $X_i$, where $i = 400$ time units (for ease we shall refer to the these time units as days). As with real-world data, this time series may contain a stochastic (noise) component ($N_i$), and a dynamically determined component ($D_i$). Figure 1a shows such a time series, where $D_i$ is represented by the addition of two oscillations of differing periods, and $N_i$ is represented by Gaussian (white) noise; the values of both $N_i$ and $D_i$ range from 0 to 1. To $X_i$, we have also added a small signal component, $S_i$, a regular 5-day deviation of 0.3 units amplitude, repeating at time intervals of 40 days throughout the 400 day time period, so that $X_i = N_i + D_i + S_i$.

Through the successive averaging of events in the composite methodology we may isolate a low-amplitude signal component from the time series. Before doing this, it is beneficial to attempt to remove variations in $X_i$ that are unconnected to the $S_i$ component and may reduce our signal-to-noise ratio (SNR). In reality we may have limited knowledge of the properties of a potential signal component within a dataset. If we suppose the signal we are testing for has an upper-limit length shorter than 7 days, we may set a filter length of three times our maximum expected signal length (e.g., 21 days) to use as a high-pass filter. This may eliminate some noise concerns while leaving our signal unaffected (potential loss of signal from filtering, and noise reduction in composites is discussed in later sections). We then calculate $F_i$, a smooth (running mean) of our dataset with a width set by our expected signal (in our example it is 21-days). The values of $F_i$ (thick line) are shown with $X_i$ in Figure 1b.

By subtracting $F_i$ from $X_i$ we obtain $A_i$, a high-pass filtered dataset, which we shall refer to throughout this work as an anomaly. With prior knowledge of the occurrence times of the signal we may construct a composite matrix, $M_{jt}$, where $j = 1, \ldots, n$ enumerates the $n$ composite events (matrix rows), and $t$ is the time dimension (matrix columns). We present the composite matrix of $A_i$ in Figure 1c, over a $t = \pm 20$ period. In any one of the 10 events in the composite matrix it would be highly difficult to objectively determine the presence of a signal at $t_0$. However, by averaging the 10 events of the matrix together to create a composite mean,

$$C_t = \frac{1}{n}\sum_{j=0}^{n} M_{jt}, \qquad (1)$$

the noise is reduced by the square root of the number of events,

$$\Delta C_t = \frac{\sigma}{\sqrt{n}}, \qquad (2)$$

where $\Delta C_t$ indicates the standard error of the mean (SEM) at $t$ and $\sigma$ is the sample standard deviation of $C_t$. As the noise diminishes the signal has an increased chance of being identified. For each time step $t$, the mean and SEM of the matrix may be calculated using Equations 1 and 2 respectively; these data are presented in Figure 1d, and indeed show the successful isolation of a signal of approximately 0.3 amplitude centered on $t_0$. This example and all those presented in this manuscript relate only to one particular statistic, the sample mean, however, any other statistic may also be used. While the composite approach appears straightforward, inconsistencies in the design, creation, and evaluation of composites can strongly affect the conclusions of composites studies and are the focus of this paper.

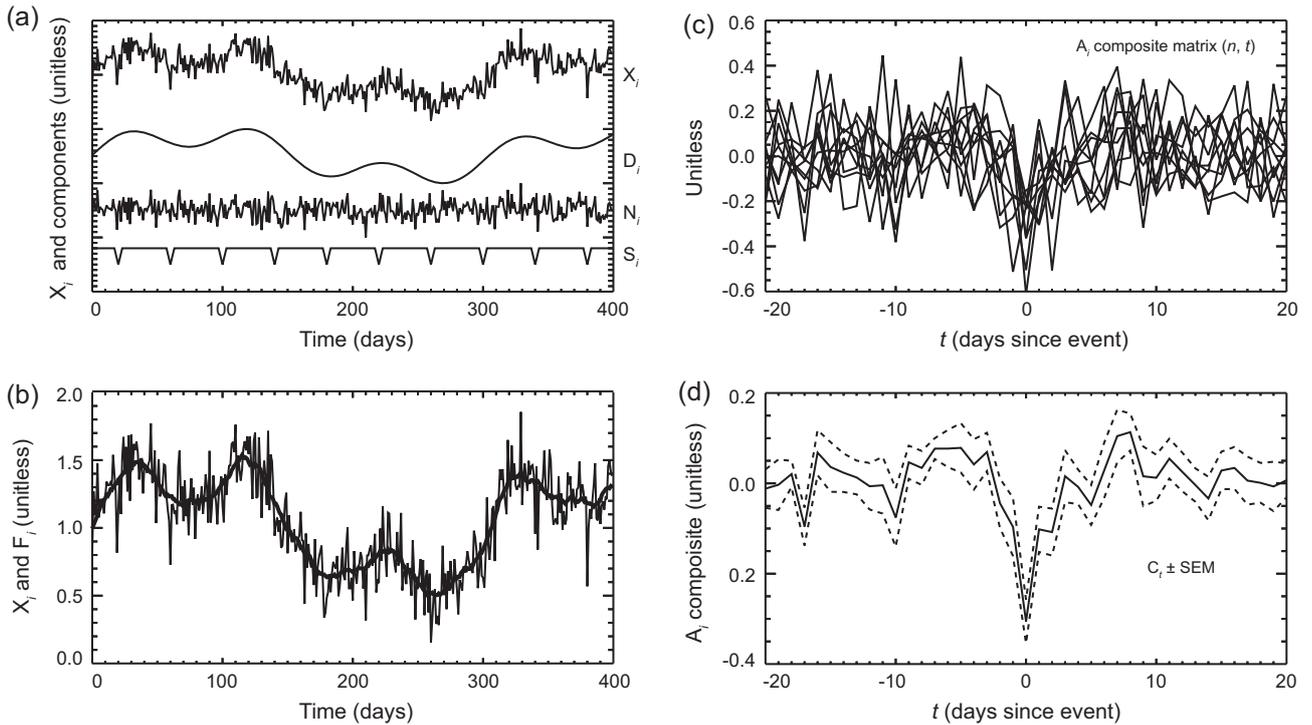

**Fig. 1.** (a) Fictional time series $X_i$, comprised of deterministic ($D_i$), stochastic ($N_i$), and a low-amplitude repeating signal ($S_i$) components. $D_i$ and $N_i$ range from 0.0 to 1.0, while $S_i$ has an amplitude of 0.3. (b) $F_i$, a 21-point smooth (box-car running mean) of $X_i$. By subtracting $F_i$ from $X_i$ a high-pass filtered dataset $A_i$ is produced. (c) A composite matrix of events from $A_i$, where rows = $n$, the number of repeating signals in $S_i$ (a composite event), and columns = $t$, the number of days since the composite event. (d) $C_t$, the composite means of $A_i$, the SEM.

Numerous composite studies have been published within the field of solar-terrestrial physics, relating to a hypothesized connection between small changes in solar activity and the Earth's atmosphere (e.g., Tinsley et al. 1989; Tinsley & Deen 1991; Pudovkin & Veretenenko 1995; Stozhkov et al. 1995; Egorova et al. 2000; Fedulina & Laštovička 2001; Todd & Kniveton 2001, 2004; Kniveton 2004; Harrison & Stephenson 2006; Bondur et al. 2008; Kristjánsson et al. 2008; Sloan & Wolfendale 2008; Troshichev et al. 2008; Svensmark et al. 2009; Dragić et al. 2011; Harrison et al. 2011; Laken & Čalogović 2011; Okike & Collier 2011; Laken et al. 2012a; Mironova et al. 2012; Svensmark et al. 2012; Artamonova & Veretenenko 2011; Dragić et al. 2013). However, despite extensive research in this area, clear conclusions regarding the validity of a solar-climate link from composite studies have yet to be drawn. Instead, the numerous composite analyses have produced widely conflicting results: some studies have shown positive statistical associations between the CR flux and cloud properties (e.g., Tinsley & Deen 1991; Pudovkin & Veretenenko 1995; Todd & Kniveton 2001, 2004; Kniveton 2004; Harrison & Stephenson 2006; Svensmark et al. 2009, 2012; Dragić et al. 2011, 2013; Harrison et al. 2011; Okike & Collier 2011), while others find no clearly significant relationships (e.g., Lam & Rodger 2002; Kristjánsson et al. 2008; Sloan & Wolfendale 2008; Laken et al. 2009; Laken & Čalogović 2011; Laken et al. 2011; Laken et al. 2012a; Čalogović et al. 2010), or even identify significant correlations of a negative sign (e.g., Wang et al. 2006; Troshichev et al. 2008). We suggest that these ambiguities may result from seemingly minor methodological differences between the composites (e.g., relating to the filtering or normalization of data), which are capable of producing widely divergent results. When the dataset is not suited to a particular method of statistical analysis, incorrect conclusions regarding the significance (i.e., the probability $p$-value associated with composite means at given times) of the composites may be reached, which is part of the reason why these aforementioned studies have presented conflicting results.

In this paper, we aim to highlight the methodologies that may produce such conflicts. We provide details on how to properly perform composite analyses of geophysical datasets, and suggest a robust method for estimating the statistical significance of variations over the composite period. Although methods to assess the significance of variations over composites that account for non-random variations have been previously demonstrated (Forbush et al. 1982, 1983; Singh 2006), the incorrect application of statistical tests remains a common feature among composite analyses. Consequently, in this work we also present a further method of assessing statistical significance, which draws from randomized samples of the datasets themselves using a Monte Carlo (MC) methodology and, as a result, makes no assumptions regarding the properties of the data. With this paper, we have also included the discussed datasets and IDL software to reproduce the methods shown here, so that our work may be readily reproduced and adapted (avaiable at www.benlaken.com/Laken_Calogovic_2013).

## 2. Working from the example of the hypothesized cosmic ray – cloud link

Throughout this work we will continue to use the example of testing the hypothesized CR flux/cloud connection to provide examples of various issues that may arise during composite analyses (albeit using real-world data rather than the idealized, fictitious data of Fig. 1). A link between the CR flux and cloud was initially suggested by Ney (1959), who theorized that the weather might be influenced by variations in the CR flux. Dickinson (1975) later proposed that such associations may result from an influence of atmospheric ionization produced by the CR flux on the microphysics of aerosols and clouds. Some of the first reports of positive correlations between clouds and cosmic radiation came from the composite studies (e.g., Tinsley et al. 1989; Pudovkin & Veretenenko 1995; Pudovkin et al. 1997). These and subsequent studies selected time periods based on the occurrence of sudden, high-magnitude reductions in the CR flux impinging on the Earth's atmosphere, termed Forbush decrease (Fd) events (Forbush 1938), generated by solar disturbances such as coronal mass ejections (Lockwood 1971; Cane 2000).

As an example, we will work from a composite of 44 Fd events identified from the Mt. Washington Observatory from 1984 to 1995, located at 44.30°N, 288.70°E, 1,900 m above mean sea level, at a cut-off rigidity of 1.24 GV (the list of Fd events were obtained from http://www.ngdc.noaa.gov/stp/solar/cosmic.html). The Fd events were adjusted so that the maximum CR flux deviation associated with each Fd is aligned with the key composite date (i.e., $t_0$ of the composite $x$-axis); without adjustment, the key date would instead be aligned to the date of Fd onset, which may differ from a period of hours-to-days from the maximal reduction in the CR flux (for further details see Laken et al. 2011). Fd events coincident within a ±7 day period of strong (>500 MeV) Solar Proton Events (SPEs) were excluded from the analysis; as such events may produce the opposite ionization effects to those we wish to isolate. Our CR flux data is the same as that of Laken et al. (2012a), being a combination of daily averaged Climax Colorado and Moscow neutron monitor data centered on zero (Fig. 2a). In addition, we have also used global daily averaged International Satellite Cloud Climatology Project (ISCCP) D1 total cloud fraction (1000–50 mb) from IR-cloudy pixels (Fig. 2b) which covers the period from 01/07/1983 to 31/12/

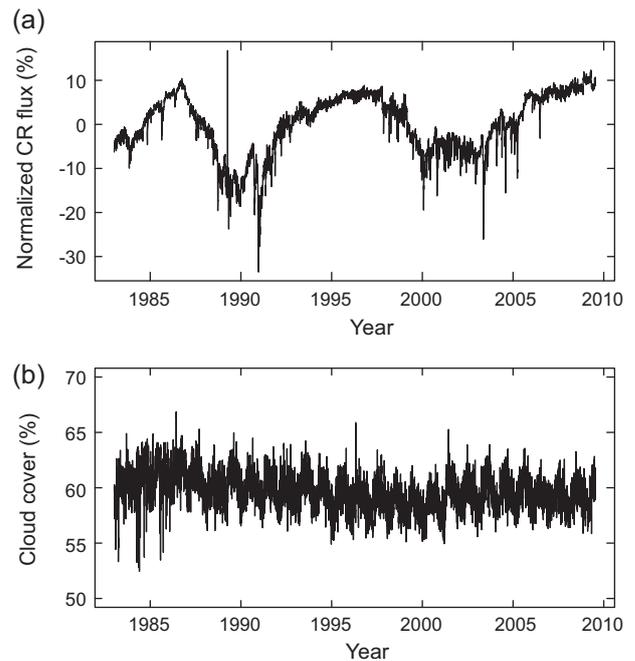

**Fig. 2.** (a) Daily averaged normalized cosmic ray flux (%) calculated from Climax Colorado and Moscow neutron monitors from Laken et al. 2012a, and (b) global, daily averaged, IR-detected cloud fractions (%) from the ISCCP D1 data.

2009. Throughout this work we will frequently use the cloud data as an anomaly equivalent to $A_i$ described in Section 1, with units of %. We again note that the analysis presented here is not meant as a serious test for the hypothesized CRcloud link, for which other similar studies exist (e.g., Laken & Čalogović 2011), but rather is presented for demonstration purposes.

## 3. Constructing a composite

### 3.1. Using composites to test a hypothesis

After successfully constructing a suitable composite for analysis, anomalies should be objectively examined for evidence of a CR flux-cloud connection via statistical methods. However, it is important to remember that statistics cannot prove any hypothesis; it can only provide a probability that a given hypothesis is or is not correct. Therefore, to test the existence of a hypothesized CR flux-cloud connection, we must construct a null hypothesis that may be tested and possibly rejected. In this instance, the null hypothesis ($H_0$) is that fluctuations observed over a composite of Fd events are indistinguishable from natural variability, while $H_1$, the alternate hypothesis, is that cloud variations distinguishable from normal (random) variability may be detected in association with the occurrence of Fd events. We must then select a confidence level at which to test our hypothesis: in this instance, we will present statistics for the commonly used 0.05 and 0.01 probability ($p$) value at the two-tailed confidence intervals (hereafter written as $p = 0.05$ and $p = 0.01$).

Detailed procedures relating to the statistical analysis of geophysical data in composite analyses have been published by Forbush et al. (1982, 1983) and Singh (2006), which demonstrate how to assess the significance of non-random (autocorrelated) data. Statistical significance has often been assessed in solar-terrestrial composite analyses by comparing composite means obtained at different times over a composite period, commonly periods prior to and following the occurrence of Fd events are utilized (e.g., Pudovkin & Veretenenko 1995; Pudovkin et al. 1997; Todd & Kniveton 2001, 2004; Svensmark et al. 2009; Dragić et al. 2011; Okike & Collier 2011). In relation to this, we importantly note that although composites of geophysical data focusing on Fd events may be considered to be independent in the $n$-dimension (events), they are often highly autocorrelated (dependent) in the $t$-dimension (time). Thus, an analysis that seeks to compare composite means across $t$ (e.g., in the manner previously described) must account for autocorrelation effects.

In autocorrelated data the value at any given point in time is affected by preceding values. As a result, sequential data points are no longer independent, and so there is less sample variation within the dataset than there would otherwise be if the points were independent. If the variance and the standard deviation ($\sigma$) are calculated using the usual formulas, they will be smaller than the true values. Instead, the number of independent data points (effective sample size) must be used to scale the sample variance (Wilks 1997). This adjustment will produce confidence intervals that are larger than when autocorrelations are ignored, making it less likely that a fluctuation in the data will be interpreted as statistically significant at any given $p$-value. In composite analyses, the number of independent data points (i.e., the effective length of the composite sequences) is equivalent to the effective sample size (Forbush et al. 1983): effective sample sizes may be calculated by the methods detailed in Ripley (2009) and Neal (1993). Despite the well-established nature of methods they are not widely applied within the solar-terrestrial community, with numerous studies implementing significance testing that simplistically assumes that the data are independent in the $t$-dimension.

In Section 3.4 we present an additional method of significance testing for composite studies based on Monte Carlo analysis approaches that is highly robust. This method does not rely on comparing composite means at given times over the composite to evaluate significance, but instead evaluates the $p$-value of each $t$-point individually, using probability density functions (PDFs) constructed from large numbers of randomly generated samples (with a unique PDF at each $t$-point).

### 3.2. Generating anomalies: the importance of static means in composite analyses

To effectively evaluate changes in daily timescale data, variations at timescales beyond the scope of the hypothesis testing should be removed from the dataset. Forbush et al. (1982, 1983) showed that bias can be introduced into traditional statistical tests if fluctuations at longer timescales are not removed. The removal of annual variations alone (which normally dominate geophysical data) will not remove fluctuations that cannot be eliminated by linear de-trending, such as those relating to the effects of planetary-to-synoptic scale systems. Such systems may extend over thousands of kilometers in area, and their influence on atmospheric variability may last from periods of days to several weeks; the random inclusion of their effects in composites may produce fluctuations at timescales shorter than seasonal considerations, yet greater than those concerning our hypothesis testing (so-called intermediate timescales) which may not be removed by linear de-trending. Consequently, when intermediate timescale fluctuations are not accounted for prior to an analysis, it may result in periods of shifted mean values and high autocorrelation being inadvertently included into composites. Suitable filters should be applied to the data to remove longer timescale variability, but to retain variations at timescales that concern the hypothesis testing. In the case of a CR flux-cloud hypothesis, a 21-day running mean (high-pass filter) may be suitable, as the microphysical effects of ionization on aerosol/cloud microphysics are expected to occur at timescales of <1 week (Arnold 2006).

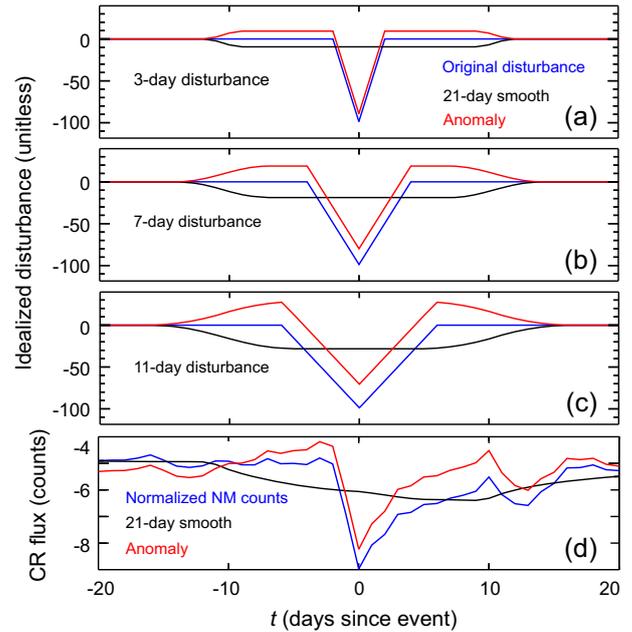

**Fig. 3.** (a–c) The differences between symmetrical idealized disturbances of increasing duration (3-day, 7-day, and 11-day) with an amplitude of 100 units (blue lines) centered on $t_0$, and the results of subtracting a 21-day smooth (box-car running mean) from the data which acts as a high-pass filter (black lines). The resulting filtered data (referred to as anomalies) are shown as red lines, and display overshoot effects. (d) The effect of subtracting the smooth filter from the CR flux (units in normalized counts). The anomaly has been offset by $-5.32$ counts in (d) for plotting purposes.

Although filtering data has the benefit of potentially reducing noise, it should be applied with caution, as it may also introduce artifacts, and reduce or even destroy potential signals altogether. Figure 3a–c shows the influence of a 21-day smooth filter on three idealized symmetrical disturbances of differing length. These disturbances are centered on $t_0$, each has an amplitude of 100 units in the $y$-axis dimension and span time periods of (a) 3-days, (b) 7-days, and (c) 11-days. To each time series, a smooth filter of 21-days (i.e., a box-car running mean) has been subtracted from the disturbance (the filters are shown as black lines). The resulting anomalies are shown as the red lines of each figure. As the duration of the idealized disturbance increases, so too does the appearance of overshoot artifacts. From Figure 3a–c, the original (peak-to-trough) signal of 0 to $-100$ has been shifted by overshoot effects by: (a) +9.6, (b) +19.0, and (c) +27.8. For a and b, the amplitude of the original disturbance is fully preserved, however, as the timescales of the disturbance increase further, the amplitude is reduced. E.g. for the 11-day disturbance the amplitude is diminished by 0.8% compared to the unfiltered signal, while for a 15-day disturbance (not shown) the amplitude decreases by 9%. The amount of signal attenuation will increase as the deviations approach the width of the smooth filter. For deviations at timescales of approximately 1/3rd the width of the smooth filter, the attenuation is negligible. Given this, filters should be applied with care to minimize noise while preserving potential signals. Following the application of the filter in this work we have constrained the reliability of our analysis to time periods of <7 days.

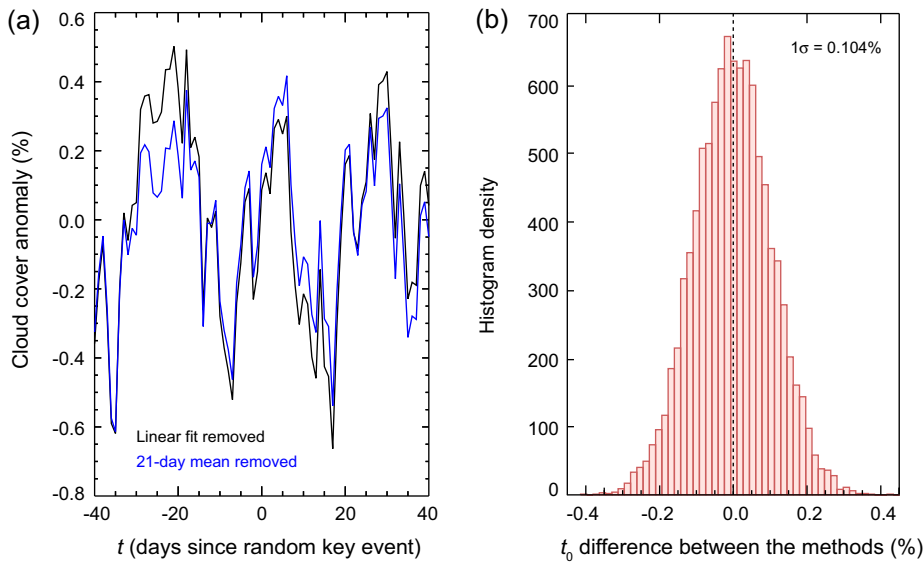

**Fig. 4.** (a) One random instance of $n = 20$ events, for both linearly detrended data (black line) and an anomaly from a 21-day moving average (blue line). Differences between these two curves show the possible influences of intermediate timescales on the data. (b) The differences between these curves at $t_0$ for 10,000 random instances of $n = 20$. The histogram shows a $1\sigma$ value of 0.1%, a non-trivial difference which may influence the outcome of a composite analysis.

Figure 3d shows a composite of $n = 44$ events, centered on the CR flux minimum of Fd events. CR flux data (blue line) are from the combined daily mean counts of the neutron monitors (NM) of Moscow and Climax, normalized by subtracting the mean of the 1950–2010 period (as in Fig. 2a). A 21-day smooth of these data are shown (black line). The normalized data show a peak-to-trough reduction between $t_{-3}$ and $t_0$ of 4.17% and a 21-day smooth (high-pass filter) of these data are also shown (black line). Following the removal of a 21-day high-pass filter from the normalized data, the resulting anomaly (red line) shows a slightly smaller peak-to-trough change over the same period of 4.05%. This indicates that 97.1% of the signal has been preserved following the application of the 21-day (high-pass) smooth filter. Overshoot distortion effects are also observable in the CR flux anomaly, and notably this produced an artificial enhancement of the increase in the CR flux prior to and following the Fd events.

Despite the limitations associated with data filtering, the noise reduction which may be gained by minimizing the influence of autocorrelated structures within data may be non-negligible. To exemplify the potential impacts of intermediate timescale structures on composites and the benefit of removing them, we have generated a random composite of cloud cover data with $n = 20$ events, over an analysis period of $t_{\pm 40}$ days. The $n = 20$ events upon which the composite was based were selected using a random number generator, ensuring that the chances of any one date (event) being selected for the composite were equal, and that once selected, a date could not re-occur in the composite. All of the random composites used throughout this work were created in this manner. The composite is presented in Figure 4a, with the cloud anomalies calculated by two methods. Firstly, through a simple removal of a linear trend from the cloud data (black line); this approach will remove linear bias from the composite resulting from seasonality. Secondly, we have also created an anomaly by subtracting a 21-day running mean (high-pass filter) from cloud cover (blue line). In this latter method, all variations (linear and non-linear) at timescales greater than 21-days are accounted for, including both seasonality and influences from intermediate timescale fluctuations. A period of 21-days was selected as this is three times greater than the upper limit response time predicted for a CR flux cloud relationship (Arnold, 2006), and so subtracting a running mean of this length from the data should not diminish the amplitude of any potential signals, yet it may still be useful in minimizing influences from longer term variations which are not the subject of the hypothesis testing. A comparison of these two approaches (Fig. 4a) gives an example of how the failure to remove non-linear intermediate timescale variations influences the composite, with noticeable deviations between the two methods. However, it is difficult to objectively assess the bias expressed as the difference between these two methods with only one random example. With this in mind histogram Figure 4b is presented, displaying the results of 100,000 randomly generated composites, showing the difference between the linear fit removed data and the 21-day smooth cloud anomaly at $t_0$. The histogram shows that the $1\sigma$ difference between these anomalies (i.e., the impact of not accounting for intermediate timescale variations) is approximately 0.1%. We note this value will vary depending on the size of $n$ and the length of the running mean applied, and some differences between the two filtering methods may also be attributed to overshoot errors. As we shall demonstrate in later sections, unaccounted-for biases of such amplitudes may have a non-negligible impact on the estimated statistical significance of composited anomalies.

### 3.3. Using signal-to-noise-ratio as a constraint when designing a composite

The data we have presented in Figure 2b are global in extent and utilize all pressure levels of the ISCCP dataset (1000–50 mb). However, composite studies often use subsets of data for analysis, with varying spatio-temporal restrictions: e.g., focusing on stratospheric layers over Antarctica (Todd & Kniveton 2001, 2004; Laken & Kniveton 2011); areas over isolated Southern Hemisphere oceanic environments at various tropospheric levels (Kristjánsson et al. 2008); low clouds over subtropicaltropical oceans (Svensmark et al. 2009); indicators of the relative intensity of pressure systems (vorticty area index) over regions of the Northern Hemisphere Mid-latitude zone, during differing seasons (Tinsley et al. 1989; Tinsley & Deen 1991); total ozone at latitudes between 40°N and 50°N, during periods of high solar activity and eastern phases of the quasi-biennial oscillation (Fedulina & Laštovička 2001); variations in atmospheric pressure and formation of cyclones/anticyclones over the North Atlantic (Artamonova & Veretenenko 2011); and diurnal temperature ranges over areas of Europe (Dragić et al. 2011; Dragić et al. 2013; Laken et al. 2012a). In addition to spatial/seasonal restrictions, it has been frequently proposed that perhaps a CR-cloud effect is only readily detectable during the strongest Fd events (e.g., Harrison & Stephenson 2006; Svensmark et al. 2009, 2012; Dragić et al. 2011). Although such propositions are rational, high-magnitude Fd events are quite rare. For example, of the 55 original Fd events between 1984 and 1995 listed in the NOAA resource (discussed in Sect. 2), the Fd intensity ($I_{Fd}$, the daily absolute percentage change of background cosmic ray intensity as observed by the Mt. Washington neutron monitor) is as follows: 28 weak events ($I_{Fd} < 5\%$), 21 moderate events ($5\% \leq I_{Fd} < 10\%$), and 6 strong events ($I_{Fd} \geq 10\%$). Consequently, composites focusing only on intense Fd events are invariably small in size, and therefore highly restricted by noise (Laken et al. 2009; Čalogović et al. 2010; Harrison & Ambaum 2010).

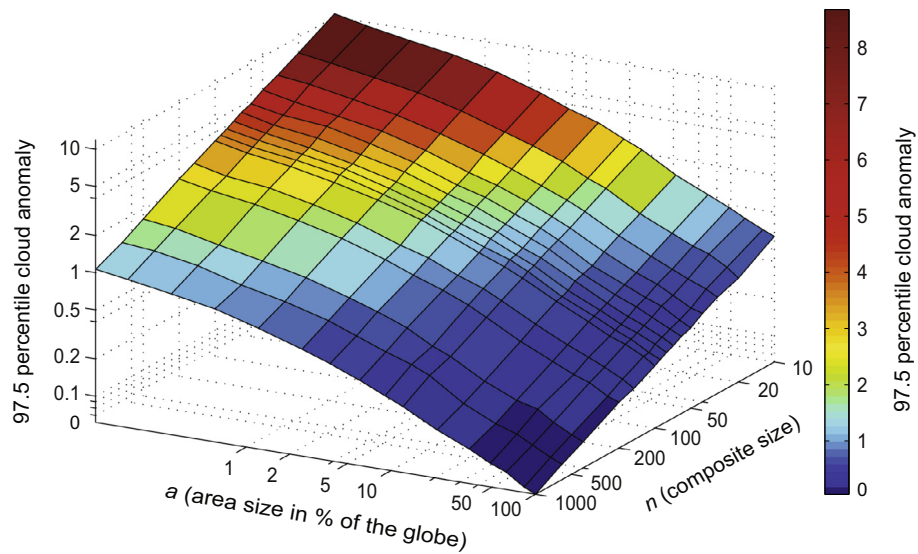

**Fig. 5.** The relationship between increasing sample area (as a percentage of global area: x-axis, denoted by $a$) and number of events in a composite (z-axis, denoted by $n$) on noise levels indicated by the 97.5 percentile mean cloud fraction anomalies (%, y-axis) within composites. These values constitute the upper $p = 0.05$ confidence interval. Each data point is calculated from 10,000 Monte Carlo (MC) simulations from cloud cover anomaly data (21-day running mean subtracted) using PDFs. All axes shown are logarithmic.

While there is often a good theoretical basis for restricting the sampling area, such constraints considerably alter the potential detectability of a signal. Restricting either the spatial area (hereafter $a$) of the sampling region or the number of sample events (composite size, hereafter $n$) will increase the background variability (noise) in composites. This relationship is quantitatively demonstrated in Figure 5, which shows how noise levels in random composites constructed from cloud cover anomalies (21-day running mean subtracted) vary for differing area $a$ and composite size $n$. Using MC simulations this is calculated for composites varying over an $a$ of 0.1–100% of the global surface, and $n$ of 10–1,000. For each combination of variables $a$ and $n$ a probability density function (PDF) of 10,000 randomly generated composite means at $t_0$ are generated and the 97.5th percentile of the distribution is then presented in Figure 5. For each corresponding $a$ and $n$, any potential CR flux-induced cloud variations (which we shall refer to as a signal) must overcome the values shown in Figure 5 in order to be reliably detected; so these values represent the upper confidence level. A nearly symmetrical 2.5 percentile lower confidence level (not shown) is also calculated for every $a$ and $n$ combination, and together these upper and lower confidence intervals constitute the $p = 0.05$ significance threshold.

The noise associated with composites of differing $a$ and $n$ sizes defines the lower-limit relationship which can be detected at a specified confidence level. Thus, for a specific magnitude of CR flux reduction, the minimum necessary efficiency at which a CR-cloud relationship must operate in order to be detected at a specified $p$-value can be calculated. For example, if composites were constructed with $n = 10$ samples from data with an area of only 1% of the Earth's surface, then a change in cloud cover would have to exceed approximately ±6.3% in order to be classified as statistically significant at the $p = 0.05$ two-tailed level.

Based on such calculations, composites can be constructed from which a response can realistically be identified. For example, the most favorable study of a CR-cloud link by Svensmark et al. (2009) finds using a sample of $n = 5$ Fd events that following an 18% reduction in the CR flux a 1.7% decrease in cloud cover occurs. Regardless of the criticisms of this study (given in Laken et al. 2009 and Čalogović et al. 2010), we use this observation to calculate the most favorable upper limit (UL) sensitivity of cloud cover changes to the CR flux (1.7/18 giving a factor of 0.094). From a consideration of this UL sensitivity and the relationship between composite noise to area $a$ and composite size $n$ in Figure 5, we may then surmise that a CR flux-induced change in cloud cover (a signal), could almost certainly never be distinguishable from background variability (noise) at the $p = 0.05$ level, from a composite of 10 Fd events with an average CR flux reduction of 3% (the value of 3% is typically used to define a Fd event, Pudovkin & Veretenenko 1995; Todd & Kniveton 2001). This is because the UL sensitivity suggests that the CR flux would at most produce a cloud signal of 0.28% ($0.094 \times 3$) assuming a linear CR-cloud relationship. Even if 100% of the globe were considered in the analysis the sample noise would still be too great (by a factor of 2) to detect the signal. For composite size of $n = 10$ Fd events we see the upper $p = 0.05$ confidence interval never drops below 0.6% due to the high noise levels of the data (Fig. 5). To significantly identify a signal with an amplitude of 0.28% in the presented cloud cover data above the $p = 0.05$ level, considering 100% of the globe, would require a composite size of at least $n = 50$.

### 3.4. Estimating the significance of anomalies by Monte Carlo methods

Evaluating the statistical significance of composited geophysical data by a comparison to randomly obtained samples is not a novel idea. One of the first applications of this method within the field of solar-terrestrial physics appears in (Schuurmans & Oort 1969) (hereafter SO69), who investigated the impacts of solar flares on Northern Hemisphere tropospheric pressures over a composite of 81 flare events. They constructed their composite over a three-dimensional (longitude, latitude, height) grid and examined changes over a 48 h period. They then evaluated the significance of their observed pressure variations at the $p = 0.05$ significance level (calculated from the standard deviation of their data). SO69 noted that due to autocorrelation in their data, the number of significant grid points beyond the $p = 0.05$ level may exceed the false discovery rate without there necessarily being a significant solar-related signal in their data (a point which we shall discuss further later in this work).

To assess the potential impact of autocorrelation on their data and more accurately gauge how unusual their result was, SO69 constructed three random composites of equal size to their original composite ($n = 81$), and compared the random results to their original findings. Although SO69 were limited by the lack of computing power available to them, and consequently their number of random composites was not sufficient to precisely identify the statistical significance of their flare composite, our method, which we will shortly describe in detail, is based upon the same principles as those of SO69. In essence, we will use populations of randomly generated composites and identify threshold significance values using probability density functions (PDFs) of these values, from which we may precisely evaluate the statistical significance of variations in the composite mean over $t$.

To achieve this, we can use the MC-based method of estimating noise used to calculate Figure 5 to provide a simple, yet powerful method of evaluating the statistical significance of composite means. From MC-generated PDFs we may test how likely it is that we can randomly obtain composite means of a given magnitude by chance. Events are selected at random from the data being scrutinized, and composites of equal sample sizes to the original composite are constructed. The data of the MC must be treated in exactly the same manner as that of the composite being evaluated with respect to the calculation of anomalies, or application of any normalization procedures (i.e., if the composite is based on an anomaly calculated from a 21-day running mean, then the MC-generated composites must also be based on data with this treatment). Following this, the composite mean at each $t$ is calculated. If enough available data exist, this procedure can be repeated many times, making it possible to build up large PDFs of composite means for each $t$. From these, confidence intervals at specific $p$-values may then be identified as percentiles of the calculated distributions. We note that this procedure may fail in cases where insufficient data exists to build up distributions which are representative of the parent population from which the samples are drawn.

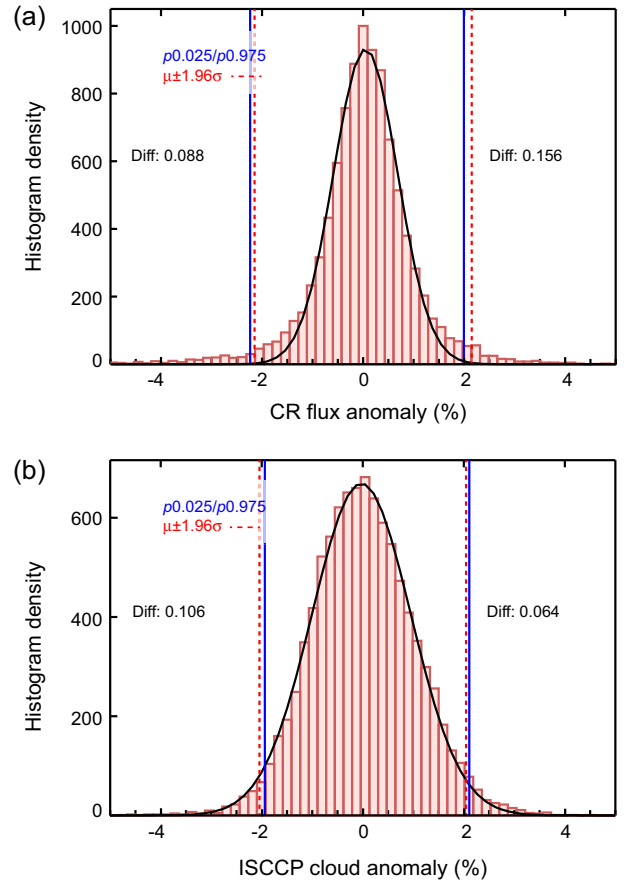

**Fig. 6.** Distribution of daily anomalies for the entire 1983–2010 data period for: (a) cosmic ray flux (%), and; (b) cloud cover (%) data. For comparison an idealized Gaussian distribution is shown on the black lines. The mean $\pm 1.96\sigma$, and 2.5/97.5th percentile values are displayed on the red dashed and blue solid lines respectively.

Geophysical data do not often follow idealized Gaussian distributions, as a result the two-tailed confidence intervals should be identified asymmetrically for optimum precision: i.e., the upper/lower $p = 0.05$ confidence interval should correspond to the 2.5th and 97.5th percentiles of the cumulative frequency, not simply to the $\pm 1.96\sigma$ value. This is demonstrated in Figure 6, which shows a distribution of both the CR flux anomaly and cloud anomaly data, comprised of all data points from 1983–2010 (in this instance the data presented are the entire population of daily values rather than composite values). The two-tailed $p = 0.05$ thresholds are calculated from both percentiles (blue lines) and from the $\pm 1.96\sigma$ level around the mean (red dashed lines). Under an ideal Gaussian distribution (displayed on the black lines) the 2.5/97.5th percentile and $\pm 1.96\sigma$ values would be both equivalent and symmetrical around the mean value. However, due to the departures from the ideal Gaussian distribution there are differences between these significance values: this is most prominent in the upper confidence interval of the CR flux data where the deviation between the $+1.96\sigma$ and 97.5th percentile values is 0.16% (Fig. 6a).

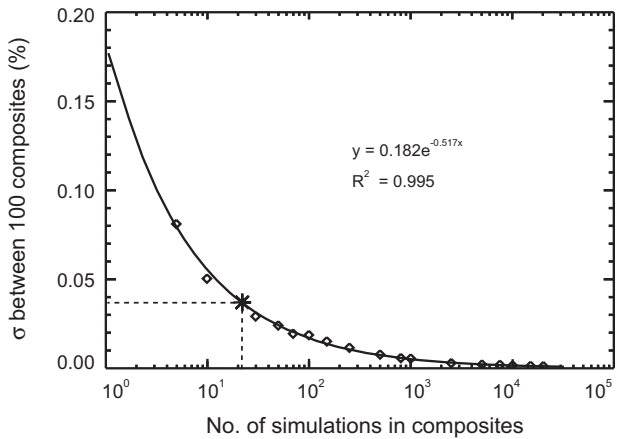

**Fig. 7.** Convergence of 100 different Monte Carlo (MC) simulations on a consistent solution, indicated by an exponential reduction in the standard deviation ($\sigma$) between the 100 means calculated for random composites of $n = 40$ events. The number of simulations within each MC increases along the $x$-axis, from 5 to 200,000 (calculated values are indicated by diamond markers). An exponential fit to these data are plotted (solid line), along with the value of the Pareto principle (i.e., the 80:20 rule), calculated from the fit, which indicates how many simulations are required for 80% of convergence to occur denoted by the asterisk marker on the fit, along with the corresponding $x$ and $y$-axis positions.

Although the deviations between the $\pm1.96\sigma$ and 97.5th percentiles presented here are relatively small, these differences can become highly exaggerated depending on the distribution of the data under investigation, and so this should be taken into proper consideration. We also stress that when using this technique to determine significance thresholds of composites, the confidence intervals should be calculated for each time step ($t$) in the composite individually. Similar solutions across the $t$-dimension will only be produced by MCs if the analyzed data possess an equal variance and static means.

When using MC methods there is a question as to how many simulations are sufficient to obtain a reliable result; if too few are used a reliable solution will not be identified (i.e., running the MC numerous times with an identical set-up will produce a wide range of values). Conversely, setting the number of simulations too high may be inconvenient as it may require considerable computation time. As the number of simulations increases in a MC it becomes increasingly more likely that by repeating the MC you would arrive at a consistent result (this is termed convergence). In general, the amount of convergence achieved by the MCs increases exponentially with increasing number of simulations. To exemplify this, we have calculated results for 100 different MCs run with identical inputs for a range of simulation sizes between 5 and 200,000. Each MC generated random composites of $n = 40$ events from cloud anomaly data, and calculated a distribution of the means at each simulation size. The $\sigma$ of the 100 means for each simulation size is plotted in Figure 7, and indeed shows an exponential reduction (convergence) with increasing simulation size. Following the 80:20 rule (also known as the Pareto principle), we find that 80% of the convergence in the case of Figure 7 occurs after only a relatively small number of simulations (only 22); the remaining precision is achieved at an exponentially decreasing rate. However, we note that as computing these values requires only relatively limited resources in this instance, it was possible for us to use large (10,000 simulations) to achieve a near-fully converged result throughout the remainder of this work.

To re-state and emphasize this procedure, in this work 10,000 MC simulations are used to generate PDFs of composite means expected at each instance of $t$, against which the significance of a composite may be evaluated. In each instance the mean of the PDFs should be zero as the 21-day filtered data has a mean close to zero (as indicated in Fig. 7), and the distribution of the PDF around zero will reflect the remaining (short-term) variability in the data. The PDFs show the variations expected from composites of a specified $n$-size in the absence of deterministic effects (i.e., when there are only random fluctuations occurring), and they therefore provide a basis from which to accept or reject the $H_0$ and $H_1$ hypothesis.

### 3.5. Applying the MC significance testing to real data

We present an example of the MC-based significance testing methodology applied to composites of Fd events for daily mean anomalies of CR flux (Fig. 8a) and cloud data (Fig. 8b). The composite of Figure 8 uses $n = 44$ adjusted Fd events, and is presented over a period of $t_{\pm40}$. Only Fd events not coincident within a $t_{\pm7}$ day period of ground level enhancement (GLE) events have been included in the composite (as described in Sect. 2). The mean of these data along with the $\pm1.96$ SEM values are shown on the solid black and dashed blue lines respectively. To these data, we have applied the previously discussed MC-based method of calculating confidence intervals as a test of statistical significance. These confidence intervals are calculated from PDFs of 10,000 MC simulations at each $t$-point, and are plotted for the $p = 0.05$ and $p = 0.01$ two-tailed confidence intervals (dashed and dotted red lines, respectively). The small variations in the individual confidence intervals of Figure 8 which can be observed across $t$ indicate the amount of convergence remaining to be achieved by the MCs.

Daily mean CR flux anomalies and the $\pm1.96$ SEM values are presented in Figure 8a: at $t_0$, CR flux anomalies of $-3.01 \pm 0.53\%$ are observed, corresponding to deviations of $18.0 \pm 3.2\sigma$ (the associated statistical significance of the mean and SEM ranges are all $p < 0.00$). Additionally, highly significant positive CR flux anomalies occur both before and after $t_0$, the largest of which being during $t_{-3}$, where a CR flux anomaly of $0.89 \pm 0.37\%$ was observed ($6.3 \pm 2.1\sigma$, again with the mean and SEM ranges all significant at $p < 0.00$). Increases of CR flux prior to the Fd correspond to the influence of a shock/sheath structure of coronal mass ejections (CME) that generate the Fd event, where the CRs are swept up and deflected by the propagating magnetic disturbance generated by a CME. Increases of CR flux after Fd events can be connected with overrecovery effects in some cases (Dumbović et al. 2012). However, in our case, the positive CR deviations are additionally influenced by the artificial overshoot effects resulting from the application of the 21-day smooth (high-pass) filter, as demonstrated in Figure 3d. Significant increases in the CR flux are also evident at $t_{+10}$ and $t_{+23}$ which are due to the unintentional inclusion of GLEs in the composite (which were only filtered within a $\pm7$ day period around $t_0$).

Cloud anomalies (Fig. 8b) showed no clear response over the composite period: the largest negative/positive deviations occurred prior to the statistically significant CR flux variations, at $t_{-26}$ ($-0.39 \pm 0.38\%$) and $t_{-18}$ ($0.33 \pm 0.32\%$), corresponding to deviations of $2.5 \pm 2.4\sigma$ and $2.1 \pm 2.1\sigma$ respectively. The $p$-value associated with these mean anomalies is $p = 0.01$ (with upper/lower SEM values of $p = 0.91/p < 0.00$) and $p = 0.04$ ($p < 0.00/p = 0.96$) respectively. All other variations shown in Figure 8, including those observed following the Fd induced CR flux reduction, were of a smaller magnitude. The second largest negative anomaly observed occurred at $t_{+5}$ and warrants discussion (for reasons which will become apparent in the next paragraph): the anomaly possesses a mean of $-0.37 \pm 0.41\%$, corresponding to a deviation of $2.4 \pm 2.6\sigma$ and a significance of $p = 0.02$ (with upper/lower SEM values of $p = 0.84/p < 0.00$).

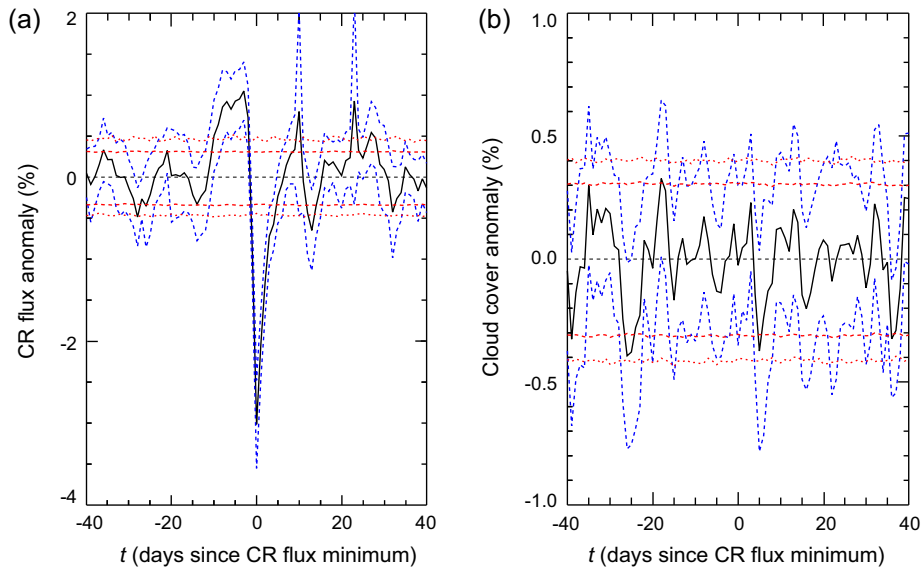

**Fig. 8.** Composites of ($n = 44$) Forbush decrease events over a $t_{\pm40}$ day period not coincident within a $t_{\pm7}$ day period of GLEs from 1983 to 1995 for: a) the CR flux (%), and b) cloud cover (%). The mean anomalies (solid black lines) are plotted together with the $\pm1.96$ SEM values (blue dashed lines) for each of the $t_{\pm40}$ days of the composite. Based on PDFs of 10 000 MC simulations at each $t$ value two-tailed confidence intervals are calculated at the $p = 0.05$ and $p = 0.01$ levels (dashed and dotted red lines respectively).

It should be stressed that by analyzing 81 points ($t_{\pm40}$) over a composite using the $p = 0.05$ significance level, there will be a false discovery rate (FDR) of approximately 4 days ($81 \times 0.05 = 4.05$) and we can expect that the mean anomaly will exceed the $p = 0.05$ level by chance on approximately four occasions over an 81-day period. Indeed, in Figure 8b we observe mean anomalies of $p < 0.05$ in six instances over the composite, in line with the expected FDR. In contrast, the CR flux is observed to be significant at the $p = 0.05$ level at 26 days out of 81 as a result of the influence of Fd and GLE events. We stress that the FDR noted here relates to the frequency with which the mean anomaly will appear as significant and not the error around the mean anomaly (SEM). These error intervals of course relate to how accurately the mean value is known, so in our case we have presented intervals which show with a $\pm95\%$ accuracy the range in which the mean value is likely to occur. Consequently, the SEM ranges are regularly seen to pass the $p = 0.05$ threshold. Evaluating the statistical significance of both the mean and its error range will give a more robust indication of statistical significance than can be obtained from the mean value alone. For example, the $t_{-3}$ and $t_0$ CR flux anomalies are unambiguously statistically significant (SEM are $p < 0.00$), however, although the $t_{-26}$ cloud anomalies show a mean and lower SEM which are statistically significant ($p < 0.05$), almost all of the upper SEM values exist at $p > 0.05$, indicating that this is not a robust result, as should be expected from a data point with its significance attributed to the FDR. Considered together these results clearly show that there are no unusual variations during or following statistically significant variations in the CR flux, and consequently support the rejection of $H_1$ and the acceptance of $H_0$.

From a combination of the observed reduction in CR flux of 3%, and the observation that no clear cloud changes occurred above the $p = 0.01$ significance level which are equivalent to cloud anomalies of about 0.40% (see Fig. 8b), we may also conclude that if a CR-cloud relationship exists, then a 1% CR flux change is, at most, able to alter cloud by $\leq 0.13\%$ (0.4%/ 3%). If a CR flux-cloud relationship were more efficient than this limit, we would be able to detect a statistically significant cloud response over daily timescales. Since we do not, our conclusions must be limited by the statistical noise of the dependent (cloud) dataset. However, we note that supporting lines of evidence suggest at least that the higher range of values associated with this upper-limit constraint is likely too large, as it implies that over a solar cycle decadal oscillations in the CR flux on the order of 20% may induce cloud changes of $\leq 2.6\%$ ($20\% \times 0.13\%$, assuming a linear CR-cloud relationship), but no such periodic variations in cloud have been identified in either ISCCP or MODIS cloud data at any atmospheric level over the past 26 years of satellite observations (Kristjánsson et al. 2004; Agee et al. 2012; Laken et al. 2012b, 2012c).

Although the MC methods as shown in the previous examples have many advantages over classical statistical tests (like the Student's $t$-test), we reiterate that there are situations where MC methods could yield incorrect or imprecise estimates of significance levels. Specifically, this may occur in instances where: (1) there is a limited amount of data to generate unique random samples. The total number of unique samples which may be generated can be calculated as

$$\text{MC}_{\text{sims}} = \frac{m!}{n!(m-n)!}. \tag{3}$$

where $\text{MC}_{\text{sims}}$ is the number of unique simulations, $n$ is the subsample size (i.e., size of composites), and $m$ is the size of the parent population. If the number of unique samples is smaller than the number of simulations required for the MC to converge, then the accuracy of the MC-analysis method will be limited, (2) the number of MC simulations is quite low, resulting in a higher uncertainty associated with any calculated confidence intervals as the MCs had not fully converged (a cause of this may be either from point 1 or by experimental design); or (3) and the analyzed data contains outliers. This would be a problem only if point 1 and/or 2 were also occurring, thereby preventing the resulting MC distributions from becoming accurate representations of their parent population.

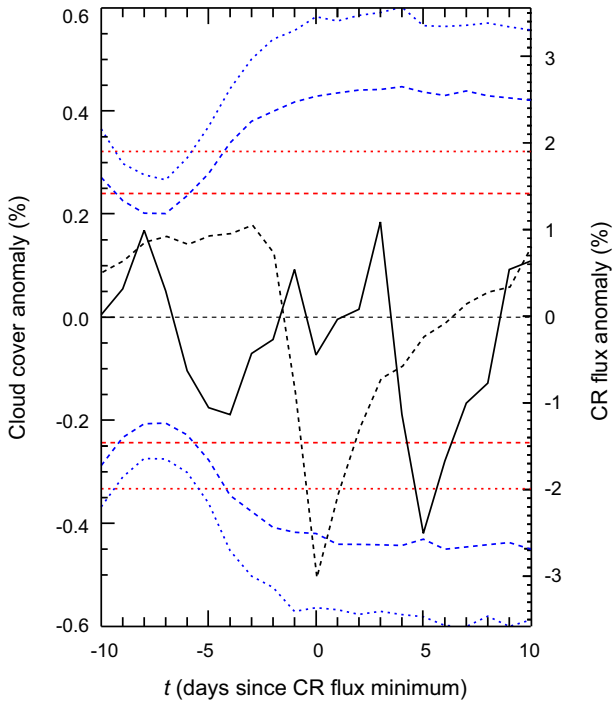

**Fig. 9.** Composite means of cloud cover anomalies (%, black solid line, left-hand axis) calculated by subtracting from the observed record the time average of the record over a five-day period starting at $t_{-10}$ (normalization period) and the corresponding CR flux anomalies (%, black dashed line, right-hand axis) processed in the same way. Based on MC simulations two sets of confidence intervals at the two-tailed $p = 0.05$ (dashed lines) and $p = 0.01$ (dotted lines) are calculated. The confidence intervals shown as red lines are calculated from a PDF of 10,000 MC simulations during the normalization period, while the confidence intervals shown as blue lines are calculated from PDFs of 10,000 MC simulations at each $t$ individually (blue lines). Red lines indicate that the $t_{+5}$ cloud anomalies are 3.5$\sigma$ ($p < 0.00$), while the lines show the same anomalies to be 1.89$\sigma$ ($p = 0.06$).

### 3.6. Common causes of false-positives

The data of Figure 8b clearly show no significant cloud response, and give no cause to support a hypothesized relationship between the CR flux and cloud cover. However, by constructing the composite in a subtly altered, yet still seemingly logical manner, highly different results can be produced. For example, if we seek to test how cloud cover varies following Fd events, it is reasonable to propose that we should compare cloud properties before Fd events to those following the events. Based on this, we may construct cloud anomalies against a supposedly undisturbed normalization period, prior to the occurrence of the Fd events, and then evaluate the significance of the changes by comparing the post-Fd event values to the values of the normalization period. While these propositions are reasonable, and this procedure has appeared in numerous publications (e.g., Kniveton 2004; Svensmark et al. 2009, 2012; Dragić et al. 2011, 2013), this approach can have a considerable impact on the statistical significance of the anomalies, as we will now demonstrate.

In Figure 9 we present cloud cover anomalies (%) calculated from the original ISCCP data (not adjusted for seasonality or mid-to-long-term variations), where the anomalies are defined in this instance by subtracting from the observed record the time average of the record over a five-day period starting at $t_{-10}$ (hereafter this five-day period is referred to as the normalization period). We note that these anomalies do not include the use of a 21-day smooth filter as in our previous examples. This different approach consequently produces values which differ from those presented in Figure 8b. Confidence intervals calculated from the normalization period alone and extended over the entire composite period (as opposed to calculating the confidence intervals at each $t$ point individually) based on 10,000 MC simulations at the $p = 0.05$ and $p = 0.01$ two-tailed levels are also presented in Figure 8 (red dashed and dotted lines respectively). In this procedure the time average of a five-day normalization period beginning at $t_{-10}$ has been subtracted from the observed record for 10,000 randomly generated composites, and the distribution of values within the normalization period has been accumulated to produce a PDF from which the confidence intervals were calculated. Relative to the normalization period of undisturbed conditions, we observe a reduction in cloud cover of −0.42% centered on $t_{+5}$ (a 3.5$\sigma$ deviation with a statistical significance of $p < 0.01$). A traditional statistical test also indicates that this anomaly is highly significant: when we compare the cloud anomalies of the normalization period against the anomalies of $t_{+5}$ using a Student's $t$-test, we obtain a $T$-statistic of 2.87, corresponding to a $p$-value of 0.004.

However, the statistical significance of this result is incorrect. A comparison between the Fd composite values over $t$ and the distribution of MC-generated composites only has meaning if the same statistic is compared. If the correct statistical methods are applied and confidence intervals are calculated for each $t$ point (blue dashed and dotted lines in Fig. 9), the $t_{+5}$ anomalies are found to have a $p$-value of 0.06, similar to the value obtained in Figure 8b. The false positive previously identified was a consequence of the violated assumption that variations over the composite are random and non-sequential (i.e., no autocorrelation). This assumption is generally untrue of geophysical data, and in the case of Figure 9, this is further exacerbated by the subtraction of the normalization period from the observed record.

Due to the temporal development of the weather, it is logical to state that the cloud conditions of today are more likely to be similar to those of yesterday than they are of last week. Consequently, any statistical tests, which compare the values of a normalization period against a possible peak (e.g., $t_{+5}$) or any subsequent values, are inherently biased and will frequently produce false-positive results (this is the reason the $t$-test gave a false-positive result). In order to correctly assess statistical significance over $t$ in a composite where a normalization period has been subtracted from the data, confidence intervals must be calculated at each $t$ individually (e.g., the blue lines of Fig. 9). For each $t$ the confidence intervals are drawn from a PDF of 10,000 MC simulations, which have been treated in an identical manner to the Fd composite (i.e., the randomly generated composites have had a five-day normalization period commencing at $t_{-10}$ subtracted from the observed mean at each $t$). From the resulting confidence intervals we clearly demonstrate that the variations of $t_{+5}$ are non-significant ($p > 0.05$).

We note that methodological differences in generating the anomalies between Figures 8 and 9 (black lines in both figures) have resulted in the $t_{+5}$ decrease being exaggerated by 0.05% in the later figure. Despite this, the anomaly appeared to be weakly significant ($p = 0.02$) in Figure 8 but not in Figure 9 ($p = 0.06$). This is because the autocorrelation effects present in the analysis of Figure 9 have also resulted in relatively wider confidence intervals. This effect is also combined with the different calculation of the $t_{+5}$ anomalies, and as a result we obtain two differing $p$-values. So which is correct? This is an intentionally misleading question, as the two results are merely independent statements of the probability of obtaining values of an equal magnitude by chance when the data is treated in a certain manner, including, excluding, or ignoring various effects. When this result is interpreted objectively, examining both the mean and error, longer-term variations, and excluding as many impertinent aspects of the data as is possible (e.g., by restricting influence from autocorrelations and minimizing noise), we may readily reach the conclusion that the $t_{+5}$ anomaly is not unusual over the composite.

If we were to make the unsupported assumption that there exists a connection between the $t_{+5}$ variation in cloud and the Fd events, then we may interpret the data to suggest that an approximate change in the CR flux of 1% may lead to global cloud changes of 0.12% (0.37/3.02, where 0.37 is the absolute cloud anomaly at $t_{+5}$ in Fig. 8b). This response is even larger than our previously discussed upper limit value of 0.09% based on Svensmark et al. (2009). Objectively considered, there is no evidence yet presented for accepting that the $t_{+5}$ anomaly is more than stochastic variability, or for making the assumption of a connection between the Fd events and cloud anomalies at $t_{+5}$. However, it may always be favorably argued that the noise levels of the experiment may mask a small CR-cloud signal overwhelmed by noise. Indeed, experiments of this manner may only restrict the potential magnitude of such a signal by increasing experimental precision, but may never disprove it entirely.

To the previous point regarding autocorrelation in the data and its influence on the width and development of the confidence intervals over $t$ we add some further remarks. If the composited cloud data of Figure 9 were to be replaced by an imaginary time series absent of autocorrelation (pure white noise), and these data were treated in the same manner (with a normalization period of five days), the resulting MC-calculated confidence intervals (equivalent to the blue lines of Fig. 9) would also display some weak non-stationary characteristics similar to those of Figure 9. The confidence intervals would be relatively narrow around the normalization period and expand (to a nearly static value) outside of the normalization period. The presence of persistent noise enhances these characteristics, resulting in narrower confidence intervals during the normalization period, which then expand with increasing time from the normalization period. The duration of this expansion and the final width of the confidence intervals are related to the amount of persistent long-memory noise within the data.

We again reiterate that traditional statistical tests (such as the Student's $t$-test) may also be used to calculate significance after an adjustment in the assumed degrees of freedom (Forbush et al. 1982, 1983), which can be done by calculating the effective sample size (Neal 1993; Ripley 2009). We note that the MC-approach of the red lines in Figure 8 and the blue lines of Figure 9 is identical, and because the confidence intervals are calculated for each time-point individually, it is able to correctly account for the effects of restricted sample variability and autocorrelation present in the data and exacerbated by the normalization procedure.

### 3.7. A note on adding dimensions

The composites discussed thus far only concern area-averaged (one-dimensional) data. Such composites are usually from either point measurements (e.g., Harrison et al. 2011) or area-averaged variations (e.g., Svensmark et al. 2009) with time as the considered dimension. A limitation of this approach is that it does not provide the capacity to differentiate between small changes over large areas, or large changes over small areas: i.e., one-dimensional composites of this nature only enable an evaluation of integrated anomalies.

By considering the data at higher dimensions (i.e., over both temporal and spatial domains), differentiation between localized high-magnitude anomalies, and low-magnitude large-area anomalies is possible. However, the increased complexity also requires increased caution, as a dataset with three dimensions (e.g., latitude, longitude, and time) is essentially a series of parallel, independent hypothesis tests (see the description in the auxiliary material of Čalogović et al. 2010). To such data, the methods of one-dimensional composite analyses previously described may be applied. Following proper construction of the composite anomalies, the area over which anomalies are statistically significant may then be evaluated; this may be done by identifying a threshold $p$-value, and assessing the significance of the data points independently. Summing the number of statistically significant data points at each $t$ of the composite gives a simple integer measure, against which the normality of the anomalies may be evaluated: this is referred to as the field significance (for further details, see Storch 1982; Livezey & Chen 1983; Wilks 2006).

If autocorrelation effects and the presence of factors influencing the data are absent (i.e., for an idealized sequentially independent random dataset), the percentage of significant pixels over the integrated field should only reflect the false discovery rate (FDR) associated with the chosen $p$-value. However, as previously noted, autocorrelation effects are normally a common feature of geophysical data. Cloud data show both spatial and temporal autocorrelation, and thus the number of significant data points at any given time is likely to be greater than expected from calculations of the FDR alone. The field significance may be effectively assessed via a variety of approaches such as the Walkers test (Wilks 2006). In addition, MC-approaches similar to those previously described in this work could be readily adapted to calculate distributions of randomly generated field significance against which to calculate a $p$-value.

While integrated (one-dimensional) composites are only able to test the net amplitude of anomalies, multi-dimensional composites may identify the extent of significant anomalies by creating a series of independent parallel hypothesis tests. However, a limiting weakness of such an approach, as previously noted, is the constraint of small sample areas on the signal-to-noise ratio (Fig. 5); this makes it less likely that a low-amplitude signal may be reliably detected over a small area. Therefore, at least in the context of studies discussed in this work, we suggest that multi-dimensional studies should be used in conjunction with one-dimensional time series analyses; e.g., to demonstrate that any statistical significance identified in one-dimensional tests does not result from localized high-amplitude noise.

### 3.8. Estimating statistical significance in subpopulations

The MC-based estimation of statistical significance we have described in this work has numerous advantages over traditional significance testing methods. However, it is not without limitations as previously stated at the end of Section 3.5. To these limitations we add a further caveat: MC-methods may provide inaccurate results where composites are constructed out of subpopulations of an original (parent) dataset. In the MC experiments presented in this work, we have assumed that the chances of any data point of a time series being included in a MC are equal. This is true, as the chance of a Fd event occurring is essentially random with respect to Earths atmosphere (although to this point we note that Fd occurrence tends to cluster around the period of solar maximum). However, it is not unusual for composite samples to be further constrained, adding additional selection criteria to composite events. Consequently, the resulting composites are created from subpopulations of the parent dataset, and thus their significance may not effectively be assessed by drawing random samples from the parent dataset.

In Section 3.3 we describe several studies where composites were restricted by Fd intensity, a common analysis approach. To continue with this example, imagine we were to restrict the $n = 55$ Fd events described in Section 3.3, to the most intense ($I_{Fd} \geq 10\%$) $n = 6$ events and we wished to identify the $p$-value of the $t_0$ composite mean. To properly account for the sample restriction using a MC approach would require the creation of numerous $n = 55$ samples from the parent dataset (we shall refer to these samples as parent composites). Importantly, each of the parent composites needs to have the correct statistical properties, e.g., in this instance, a mean at $t_0$ comparable to that of the composite prior to restriction (i.e., the $n = 55$ Fd composite at $t_0$). MC populations would then need to be constructed from $n = 6$ subsamples, drawn from the parent composites. We may then use PDFs of the MC results to identify the $p$-value of the $t_0$ mean for the $I_{Fd} \geq 10\%$ composite.

This change in methodology is necessary as the hypothesis test is no longer concerned with determining the chance of obtaining a $t_0$ mean of $n = 6$ randomly, but rather, it is now concerned with determining the chance of obtaining a $t_0$ mean from $n = 6$ subpopulation when you begin with a parent composite of $n = 55$ with a specific mean value at $t_0$. i.e., the question becomes, if a group of events of $n = 55$ has a specific mean, what are the chances that a subsample of $n = 6$ of these events will have another specific mean?

## 4. Conclusions

Although numerous composite analyses have been performed to examine the hypothesized link between solar activity and climate, widely conflicting conclusions have been drawn. In this paper we have demonstrated that the cause of these conflicts may relate to differences in the various methodological approaches employed by these studies and the evaluation of the statistical significance of their results. We find that numerous issues may affect the analyses, including: (1) issues of signal-to-noise ratios connected to spatio-temporal restrictions; (2) interference from variability in data at timescales greater than those concerning hypothesis testing, which may not necessarily be removed by accounting for linear trends over the composite periods; (3) normalization procedures which affect both the magnitude of anomalies in composites, and estimations of their significance; (4) the application of statistical tests unable to account for autocorrelated data and biases imposed by the use of improper normalization procedures.

Statistical methods for correctly assessing significance in composites taking into account effective sample sizes have been previously established (Forbush et al. 1982, 1983; Singh 2006; Dunne et al. 2012). To these procedures, we add the composite construction outlined in this work, and a further robust procedure for the estimation of significance based on a Monte Carlo approach.

*Acknowledgements.* We kindly thank Beatriz González Merino (Instituto de Astrofísica de Canarias), Thierry Dudock de Wit (University of Orleans), Jefrey R. Pierce (Colorado State University), Joshua Krissansen-Totton (University of Auckland), Eimear M. Dunne (Finnish Meteorological Institute), and two anonymous referees for their helpful comments. The list of Forbush decrease and Ground Level Enhancements was obtained from http://www.ngdc.noaa.gov/stp/solar/cosmic.html. Cosmic ray data were obtained from the Solar Terrestrial physics division of IZMIRAN from http://helios.izmiran.rssi.ru. The ISCCP D1 data are available from the ISCCP Web site at http://isccp.giss.nasa.gov/, maintained by the ISCCP research group at the NASA Goddard Institute for Space Studies. This work recieved support from the European COST Action ES1005 (TOSCA).

## References

Agee, E.M., K. Kiefer, and E. Cornett, *J. Clim.*, **25 (3)**, 1057, 2012.
Arnold, F., *Space Sci. Rev.*, **125 (1)**, 169–186, 2006.
Artamonova, I., and S. Veretenenko, *J. Atmos. Sol.-Terr. Phy.*, **73 (2)**, 366–370, 2011.
Bondur, V., S. Pulinets, and G. Kim, *Doklady Earth Sciences* (Springer), **422**, 1124–1128, 2008.
Čalogović, J., C. Albert, F. Arnold, J. Beer, L. Desorgher, and E. Flueckiger, *Geophys. Res. Lett.*, **37 (3)**, L03802, 2010.
Cane, H.V., *Space Sci. Rev.*, **93 (1)**, 55–77, 2000.
Chree, C., *Philos. Trans. R. Soc. London Ser. A*, **212**, 75–116, 1913.
Chree, C., *Philos. Trans. R. Soc. London Ser. A*, 245–277, 1914.
Dickinson, R.E., *Bull. Am. Meteorol. Soc.*, **56**, 1240–1248, 1975.
Dragić, A., I. Anicin, R. Banjanac, V. Udovicic, D. Jokovic, D. Maletic, and J. Puzovic, *Astrophys. Space Sci. Trans.*, **7**, 315–318, 2011.
Dragić, A., N. Veselinović, D. Maletić, D. Joković, R. Banjanac, V. Udovičić, and I. Aničin, *Further investigations into the connection between cosmic rays and climate* [arXiv:1304.7879], 2013.
Dumbović, M., B. Vršnak, J. Čalogović, and R. Župan, *A&A*, **538**, 2012.
Dunne, E.M., L.A. Lee, C.L. Reddington, and K.S. Carslaw, *Atmos. Chem. Phys.*, **12 (23)**, 11573–11587, 2012.
Egorova, L.V., V.Y. Vovk, and O.A. Troshichev, *J. Atmos. Sol.-Terr. Phy.*, **62**, 955–966, 2000.
Fedulina, I., and J. Laštovička, *Adv. Space. Res.*, **27 (12)**, 2003–2006, 2001.
Forbush, S.E., *Phys. Rev.*, **54 (12)**, 975, 1938.
Forbush, S., S. Duggal, M. Pomerantz, and C. Tsao, *Rev. Geophys. Space Phys.*, **20 (4)**, 971976, 1982.
Forbush, S., M. Pomerantz, S. Duggal, and C. Tsao, *Sol. Phys.*, **82 (1)**, 113–122, 1983.
Harrison, R.G., and M.H. Ambaum, *J. Atmos. Sol.-Terr. Phy.*, **72 (18)**, 1408–1414, 2010.

Harrison, R.G., M.H. Ambaum, and M. Lockwood, *Proc. R. Soc. London Ser. A*, **467 (2134)**, 2777–2791, 2011.

Harrison, R.G., and D.B. Stephenson, *Proc. R. Soc. London Ser. A*, **462 (2068)**, 1221–1233, 2006.

Kniveton, D., *J. Atmos. Sol.-Terr. Phy.*, **66 (13–14)**, 1135–1142, 2004.

Kristjánsson, J., J. Kristiansen, and E. Kaas, *Adv. Space. Res.*, **34 (2)**, 407–415, 2004.

Kristjánsson, J.E., C.W. Stjern, F. Stordal, A.M. Fjsraa, G. Myhre, and K. Jonasson, *Atmos. Chem. Phys.*, **8 (24)**, 7373–7387, 2008.

Laken, B., and J. Čalogović, *Geophys. Res. Lett.*, **38 (24)**, L24811, 2011.

Laken, B., and D. Kniveton, *J. Atmos. Sol.-Terr. Phy.*, **73 (2)**, 371–376, 2011.

Laken, B., A. Wolfendale, and D. Kniveton, *Geophys. Res. Lett.*, **36**, L23803, 2009.

Laken, B., D. Kniveton, and A. Wolfendale, *J. Geophys. Res.*, **116**, D09201, 2011.

Laken, B., J. Čalogović, T. Shahbaz, and E. Palle, *J. Geophys. Res.*, 10–1029, 2012a.

Laken, B., E. Pallé, and H. Miyahara, *J. Clim.*, 2012b.

Laken, B., E. Pallé, J. Čalogović, and E. Dunne, *J. Space Weather Space Clim.*, **2 (A18)**, 13, 2012c.

Lam, M., and A. Rodger, *J. Atmos. Sol.-Terr. Phy.*, **64 (1)**, 41–45, 2002.

Livezey, R.E., and W. Chen, *Mon. Weather Rev.*, **111**, 46–59, 1983.

Lockwood, J.A., *Space Sci. Rev.*, **12 (5)**, 658–715, 1971.

Mironova, I., I. Usoskin, G. Kovaltsov, and S. Petelina, *Atmos. Chem. Phys.*, **12**, 769–778, 2012.

Neal, R.M., *Technical Report CRG-TR-93-1*, 1–144, 1993.

Ney, E.P., *Nature*, **183**, 1959.

Okike, O., and A.B. Collier, *General Assembly and Scientific Symposium, 2011 XXXth URSI IEEE*, 1–4, 2011.

Pudovkin, M., and S. Veretenenko, *J. Atmos. Terr. Phys.*, **57 (11)**, 1349–1355, 1995.

Pudovkin, M., S. Veretenenko, R. Pellinen, and E. Kyrö, *Adv. Space. Res.*, **20 (6)**, 1169–1172, 1997.

Ripley, B.D., *Stochastic simulation* (Wiley), **316**, 2009.

Schuurmans, C., and A. Oort, *Pure Appl. Geophys.*, **75 (1)**, 233–246, 1969.

Singh, Y.P., Statistical considerations in superposed epoch analysis and its applications in space research. *J. Atmos. Sol.-Terr. Phy.*, **68 (7)**, 803–813, 2006.

Sloan, T., and A.W. Wolfendale, *Environ. Res. Lett.*, **3 (2)**, 024001, 2008.

Storch, H.V., *J. Atmos. Sci.*, **39**, 187–189, 1982.

Stozhkov, Y.I., I. Martin, G. Pellegrino, H. Pinto, G. Bazilevskaya, P. Bezerra, V. Makhmutov, N. Svirzevsky, and et al., *Il Nuovo Cimento C*, **18 (3)**, 335–341, 1995.

Svensmark, H., T. Bondo, and J. Svensmark, *Geophys. Res. Lett.*, **36 (15)**, L1501, 2009.

Svensmark, J., M. Enghoff, and H. Svensmark, *Atmos. Chem. Phys. Discuss.*, **12 (2)**, 2012.

Tinsley, B.A., G.M. Brown, and P.H. Scherrer, *J. Geophys. Res.*, **94 (D12)**, 14783–14, 1989.

Tinsley, B.A., and G.W. Deen, *J. Geophys. Res.*, **96 (22)**, 283–22, 1991.

Todd, M.C., and D.R. Kniveton, *J. Geophys. Res.*, **106 (D23)**, 32031–32032, 2001.

Todd, M.C., and D.R. Kniveton, *J. Atmos. Sol.-Terr. Phy.*, **66 (13)**, 1205–1211, 2004.

Troshichev, O., V. Vovk, and L. Egorova, *J. Atmos. Sol.-Terr. Phy.*, **70 (10)**, 1289–1300, 2008.

Wang, M., L. He, and H. Jia, *High Energy Phys. Nucl. Phys.-Beijing*, **30 (1)**, 75, 2006.

Wilks, D., *J. Clim.*, **10 (1)**, 65–82, 1997.

Wilks, D., *J. Appl. Meteorol. climatolo.*, **45 (9)**, 1181–1189, 2006.